\documentclass[aps,longbibliography,pre,reprint,bibnotes]{revtex4-1}

\usepackage{amsmath,amsfonts,amssymb,amsthm}
\usepackage{mathtools}
\usepackage{xcolor}
\usepackage[colorlinks=true,linkcolor=black,citecolor=black,urlcolor=black,bookmarks,breaklinks=true]{hyperref}
\usepackage{float}

\newcommand{\nc}{\newcommand}
\nc{\I}{{\mathbf 1}}
\nc{\bN}{{\mathbf N}}
\nc{\bM}{{\mathbf M}}
\nc{\cB}{{\mathcal B}}
\nc{\cL}{{\mathcal L}}
\nc{\R}{{\mathbb R}}
\nc{\N}{{\mathbb N}}
\nc{\Z}{{\mathbb Z}}
\nc{\md}{\mathrm{d}}
\nc{\BP}{\mathbb{P}}
\nc{\BE}{\mathbb{E}}
\nc{\BQ}{\mathbb{Q}}
\nc{\COV}[1]{\mathsf{Cov}\left( #1 \right)}

\usepackage{physics, bm}
\nc{\fn}[2]{\mathinner{#1\mathopen{\left(#2\right)}}}
\nc{\vect}[1]{\bm{#1}}

\nc{\nv}[1]{\fn{\sigma_N^2}{#1}}
\nc{\E}[1]{\BE\left[#1\right]}
\nc{\En}[1]{\BE_{n}\left[#1\right]}

\nc\ch[1]{\DIFaddbegin \DIFadd{#1}\DIFaddend}
\nc\nch[1]{\textcolor{red}{#1}}
\usepackage{soul,xcolor}
\setstcolor{red}

\nc{\lattice}{\mathcal{L}}

\begin{document}

\title{Structural Characterization of Many-Particle Systems on Approach to Hyperuniform States}

\author{Salvatore Torquato}
\email[]{Email: torquato@princeton.edu}
\affiliation{Department of Chemistry, Department of Physics, Princeton Institute for the Science and Technology of Materials, and Program in Applied and Computational Mathematics, Princeton University, Princeton, New Jersey 08544, USA}
\date{\today}

\begin{abstract}

The study of hyperuniform states of matter is an emerging multidisciplinary field, impinging on 
topics in the physical sciences, mathematics and biology.
The focus of this work is the exploration of quantitative descriptors that 
herald when a many-particle system in $d$-dimensional Euclidean space $\mathbb{R}^d$ 
approaches a hyperuniform state as a function of the relevant control parameter. We establish quantitative criteria to ascertain the extent of hyperuniform and nonhyperuniform
distance-scaling regimes as well as the crossover point between them
in terms of the ``volume" coefficient $A$ and ``surface-area" coefficient
$B$ associated with  the local number variance $\sigma^2(R)$ for a spherical window of radius $R$.  The larger the ratio $B/A$, the larger 
the hyperuniform scaling regime, which becomes of infinite extent in the limit $B/A \to \infty$.
To complement the known direct-space representation
of the coefficient $B$ in terms of the total correlation function $h({\bf r})$, we derive 
its corresponding Fourier representation in terms of the structure factor $S({\bf k})$, which
is especially useful when scattering information is available experimentally or theoretically. 
We also demonstrate that the free-volume theory of the pressure of
equilibrium packings of identical hard spheres that approach  a strictly jammed state either 
along the stable crystal or metastable disordered branch  dictates that such end states be exactly hyperuniform.
Using the ratio $B/A$, as well as other diagnostic measures of hyperuniformity, including the hyperuniformity
index $H$ and the direct-correlation function length scale $\xi_c$, we study three different 
exactly solvable models as a function of the relevant control parameter, either
density or temperature, with end states that are perfectly hyperuniform.
Specifically, we analyze equilibrium  systems of hard rods and ``sticky" hard-sphere systems
in arbitrary space dimension $d$ as a function of density. We also examine low-temperature excited
states of many-particle systems interacting with ``stealthy" long-ranged pair interactions as the
temperature tends to zero, where the ground states  are disordered, hyperuniform and infinitely degenerate.
We demonstrate that our various diagnostic hyperuniformity measures are positively correlated with one another.
The same diagnostic measures can be used to detect the degree to which imperfections 
in nearly hyperuniform systems cause deviations from perfect hyperuniformity. Moreover, the capacity to identify hyperuniform scaling regimes  
should be particularly useful in analyzing experimentally- or computationally-generated
samples that are necessarily of finite size.

\end{abstract}

\maketitle

\section{Introduction}

A hyperuniform point configuration in $d$-dimensional Euclidean
space $\mathbb{R}^d$ is characterized by an anomalous suppression of
large-scale density fluctuations relative to those in typical disordered systems, such as liquids and structural glasses \cite{To03a,To18a}. More precisely, a hyperuniform point pattern is one in which
the structure factor $S({\bf k})\equiv 1+\rho\tilde{h}(\bf{k})$
tends to zero as the wavenumber $k\equiv |\bf{k}|$ tends to zero~\cite{To03a,
To18a}, i.e.,
\begin{align}
  \lim_{|{\bf k}|\to 0}S({\bf k})=0,
  \label{eq:defhyperuniformity}
\end{align}
where $\tilde{h}(\bf{k})$ is the Fourier transform of the total
correlation function $h({\bf r})\equiv g_2(\bf{r})-1$ and $g_2({\bf r})$ is
the pair-correlation function \cite{Han13}.
The hyperuniformity concept generalizes the traditional notion of long-range order in many-particle systems to include all perfect crystals, perfect quasicrystals,
and exotic amorphous states of matter. Disordered hyperuniform
materials  can have advantages over crystalline
ones, such as unique or nearly optimal, direction-independent physical properties and robustness against defects \cite{Fl09b,De15,Le16,Ma16,Zh16b,Gk17,Fr17,Ch18a,Zhang18b,Go19,Sh20,Ki20a,Yu21}.

An equivalent definition of hyperuniformity is based on the local number
variance $\sigma^2(R)\equiv \langle N(R)^2 \rangle -\langle N(R) \rangle^2$  associated with the number
$N(R)$ of points within a $d$-dimensional spherical observation window  of radius $R$, where angular brackets
denote an ensemble average.
A point pattern in $\R^d$ is hyperuniform if its variance
grows in the large-$R$ limit slower than $R^d$.
This behavior is to be   contrasted with those of  typical disordered systems, such as Poisson point
patterns, gases and liquids, where the number variance scales like the volume
$v_1(R)$ of the observation window, which is given by
\begin{equation}
v_1(R)=\frac{\pi^{d/2} R^d}{\Gamma(1+d/2)}.
\label{v1}
\end{equation}

Consider systems that are characterized by a structure factor with a radial power-law form in the
vicinity of the origin, i.e.,
\begin{align}
  S({\bf k}) \sim |{\bf k}|^\alpha \quad \text{for } |{\bf k}|\to 0.
  \label{eq:Sk-scaling}
\end{align}
For hyperuniform systems, the exponent $\alpha$ is positive ($\alpha>0$)
and its value  determines three different large-$R$ scaling behaviors  of the
number variance \cite{To03a,Za09,To18a}:
\begin{align}
  \sigma^2(R)&\sim\left\{\begin{array}{l l}
    R^{d-1},      &\alpha>1 \text{ (class I)}\\
    R^{d-1}\ln R, &\alpha=1 \text{ (class II)\,.}\\
    R^{d-\alpha}, &\alpha<1 \text{ (class III)}
  \end{array}\right. 
  \label{eq:sigma-scaling}
\end{align}
These scalings of $\sigma^2(R)$ define three classes of
hyperuniformity~\cite{To18a}, with classes I and III describing the
strongest and weakest forms of hyperuniformity, respectively.
States of matter that belong to class I include all perfect
crystals \cite{To03a,Za09}, many perfect quasicrystals \cite{Za09,Li17,Og17}, and
``randomly" perturbed crystal structures  \cite{Ga04b,Ga04,Ga08,Ki18a},
classical disordered ground states of matter \cite{To03a,Uc04b,To15} as well as 
systems out of equilibrium \cite{Zh16a,Le19a}. Class~II hyperuniform systems include some
quasicrystals~\cite{Og17}, the positions of the prime numbers \cite{To19}, 
and many disordered classical ~\cite{Do05d,Za11a,Ji11c,At16a,Zh16a} 
and quantum ~\cite{Fe56,Re67,To08b} states of matter. Examples of class~III hyperuniform systems include
classical disordered ground states~\cite{Za11b}, random organization models~\cite{He15} and
perfect glasses~\cite{Zh16a}.

By contrast, for any nonhyperuniform system, it is shown in Appendix  \ref{scalings} that
the local variance has the following large-$R$ scaling behaviors:
\begin{align}  
\sigma^2(R) \sim 
\begin{cases}
R^{d}, & \alpha =0^+ \quad \text{(typical nonhyperuniform)}\\
R^{d-\alpha}, & \alpha < 0 \quad \text{(anti-hyperuniform)}.\\
\end{cases}
\label{sigma-nonhyper}
\end{align}
For a  ``typical" nonhyperuniform system, $S(0)$ is bounded  \cite{To18a}. In {\it anti-hyperuniform} systems,
$S(0)$ is unbounded, i.e.,
\begin{equation}
\lim_{|{\bf k}| \to 0} S({\bf k})=+\infty,
\label{antihyper}
\end{equation}
and hence  are diametrically opposite to hyperuniform systems.
Anti-hyperuniform systems include fractals, systems at thermal critical points (e.g., liquid-vapor and magnetic critical points) \cite{Wi65,Ka66,Fi67,Wi74,Bi92}
as well as certain substitution tilings \cite{Og19}.

Our main concern in this paper is the exploration of quantitative descriptors that 
herald when a many-particle system is nearly hyperuniform or 
approaching a hyperuniform state, whether ordered or not. 
Elucidating such questions  not only is expected to lead to a deeper fundamental 
understanding of the nature and formation of  hyperuniform systems but has great practical value. 
For example, since hyperuniformity can endow a system with novel or optimal
physical properties, it is essential to know how
close the system must be to perfect hyperuniformity without significantly
degrading its ideal performance. 
In practice, perfect hyperuniformity is never achieved
due to defects that are inevitably present in any real finite-sized system \cite{Dr15}, whether
crystalline, quasicrystalline or disordered \cite{Ki18a}. 

We begin by recalling pertinent previous concepts and results (Sec. \ref{back}), 
including the fluctuation-compressibility theorem,
hyperuniformity as a critical phenomenon, a hyperuniformity length scale $\xi_c$
and a hyperuniformity index $H$. The latter two quantities provide  measures
of nearness to hyperuniformity. 

In the remainder of the paper, we obtain  a variety  of theoretical results
to study the problem at hand. First, for a general system, we establish quantitative criteria
to ascertain the extent of hyperuniform and nonhyperuniform
distance-scaling regimes as well as the crossover point between them
in terms of the ``volume" coefficient $A$ and ``surface-area" coefficient
$B$ associated with  the variance $\sigma^2(R)$ (Sec. \ref{scaling}). Specifically, the ratio
$B/A$ determines the crossover length scale $R_c$. The larger the ratio $B/A$, the larger 
the hyperuniform scaling regime, which becomes of infinite extent in the limit $B/A \to \infty$.
This capacity to determine hyperuniform scaling regimes  is expected to be particularly useful in analyzing experimentally- or computationally-generated samples that are necessarily of finite size.

Second, to complement the known direct-space representation
of the coefficient $B$ in terms of the total correlation function $h({\bf r})$ \cite{To03a}, we derive 
here its corresponding Fourier representation in terms of the structure factor $S({\bf k})$ (Sec. \ref{Fourier}).
The latter representation is particularly useful when the scattering
intensity is available experimentally or if the structure factor
is known analytically. 

Third, we show that the free-volume theory of the pressure of
equilibrium packings of identical hard spheres that approach
either a strictly jammed crystal or disordered state dictates
that such jammed states be perfectly hyperuniform (Sec. \ref{free}).
We describe why this outcome implies that such jammed states
must be defect-free.

Fourth, motivated by a desire to rely on analytical rather than numerical methods, we structurally characterize  three different 
disordered-system models as a function of the relevant control parameter, either
density or temperature, with end states that are perfectly hyperuniform.
These models are distinguished from most other models with hyperuniform states in that their  pair correlation functions and structure factors
are known exactly for all values of the control parameter in the thermodynamic limit. We purposely avoid the use of
simulations of many-particle systems in finite boxes  to draw conclusions, since hyperuniformity in an infinite-wavelength
property. The first model that we characterize is an equilibrium  system of hard rods 
(Sec. \ref{rods}). Here the control parameter is the number density $\rho$ (or packing fraction) and its terminal
value corresponds to the jammed state that is the integer lattice. The second model studied is a certain ``sticky" hard-sphere system
in arbitrary space dimension $d$ as a function of the number density (or packing fraction) (Sec. \ref{stick}). The third model that we characterize are low-temperature excited
states of so-called stealthy long-ranged pair interactions in $\mathbb{R}^d$ (Sec. \ref{stealthy}).
Here, the control parameter is the temperature $T$ and the corresponding ground states
at $T=0$ are disordered and degenerate.

\section{Compressibility, Inverted Critical Point,  Growing Length Scale and Hyperuniformity
Index}
\label{back}

In this section, we briefly review pertinent background material. 
This outline includes the implications of the fluctuation-compressibility theorem,
hyperuniformity as a critical phenomenon, 
a length scale that grows on approach to a hyperuniform state,
and the hyperuniformity index.

\subsection{Fluctuation-Compressibility Theorem}
\label{f-c}  

Let us now recall the well-known {\it fluctuation-compressibility theorem}
that links the isothermal compressibility $\kappa_T$ of
equilibrium  single-component many-particle ensembles at number density $\rho$ and temperature $T$
 to infinite-wavelength density fluctuations \cite{Han13}.
In particular, for  ``open" systems in equilibrium, one has
\begin{equation}
\rho k_B T \kappa_T = \frac{ \langle N^2 \rangle_{*} -  \langle N \rangle_{*}^2}{ \langle N \rangle_{*}}
= S({\bf k=  0})= 1+\rho \int_{\mathbb{R}^d} h({\bf r}) d{\bf r},
\label{comp}
\end{equation}
where $k_B$ is Boltzmann's constant and $\langle \rangle_*$ denotes an average
in the grand canonical ensemble. 

The fluctuation-compressibility relation (\ref{comp})
enables one to draw useful conclusions about the hyperuniformity of equilibrium
systems. Any ground state ($T=0)$ in which the isothermal
compressibility $\kappa_T$ is bounded and positive must be hyperuniform
because the structure  factor $S({\bf k =0})$ must be zero according to relation (\ref{comp}) \cite{To18a}.
 More generally, we infer from (\ref{comp}) that if the product
$T \kappa_T$ tends to a nonnegative constant $c$ in the limit $T \rightarrow 0$,
then the ground-state of this system in this zero-temperature limit must be nonhyperuniform if $c>0$
or  hyperuniform if $c=0$. By the same token, this means that increasing the temperature
by an arbitrarily small positive amount when a system is initially at a
hyperuniform  ground state will destroy perfect hyperuniformity,
since   $S({\bf k =0})$ must deviate from zero by some small amount determined
by the temperature dependence of the product $\kappa_T T$ for small $T$ \cite{To18a}.  This indirectly implies that phonons
or vibrational modes for sufficiently small $T$ generally destroy the hyperuniformity
of ground states \cite{To15,Ki18a}. Additionally, in order to have a hyperuniform
system that is in equilibrium at any positive $T$, the isothermal compressibility must be zero, i.e.,
the  system must be thermodynamically incompressible \cite{To18a}.



\subsection{Inverted Critical Point and Scaling Laws}
\label{inverted}

The direct correlation function $c({\bf r})$ of a hyperuniform
system behaves in an unconventional manner compared to that
of typical liquids. This function is defined via  the Ornstein-Zernike integral equation \cite{Or14}:
\begin{equation}
h({\bf r})=c({\bf r})+\rho c({\bf r}) \otimes  h({\bf r}),
\label{OZ}
\end{equation}
where $\otimes$ denotes a convolution integral.
Fourier transforming (\ref{OZ}) and solving for ${\tilde c}({\bf k})$, the Fourier transform of $c(\bf r)$, yields
\begin{equation}
{\tilde c}({\bf k})= \frac{{\tilde h}({\bf k})}{S({\bf k})}=\frac{{\tilde h}({\bf k})}{1+\rho{\tilde h}({\bf k})}.
\label{c}
\end{equation}
By definition, a hyperuniform system is one in which ${\tilde h}({\bf k=0})=-1/\rho$,
i.e., the volume integral of $h(\bf r)$ exists, implying that
$h({\bf r})$ is sufficiently short-ranged in the sense  that it decays to zero
faster than $|{\bf r}|^{-d}$. Interestingly, this
means that the denominator on the right side of
(\ref{c}) vanishes at ${\bf k=0}$
and therefore ${\tilde c}({\bf k=0})$ diverges to $-\infty$.
 This behavior implies that the the volume integral of $c({\bf r})$ does not exist
and hence the real-space direct correlation function
$c(\bf r)$ is long-ranged, i.e., decays slower than $|{\bf r}|^{-d}$.
We see that this behavior stands in diametric contrast to standard thermal critical-point systems
in which the total correlation function is long-ranged and the direct correlation function
is short-ranged such that its volume integral exists \cite{Wi65,Ka66,Fi67,Wi74}. For this reason, it has been said that
hyperuniform systems are at an ``inverted" critical point \cite{To03a}.
As noted earlier,  systems at  thermal critical points are anti-hyperuniform.

There is a class of disordered hyperuniform systems with concomitant critical exponents \cite{To03a,To18a}.
For such hyperuniform critical systems, the direct correlation function has the following asymptotic
behavior for large $r\equiv |{\bf r}|$ and sufficiently large $d$:
\begin{equation}
c({\bf r}) \sim -\frac{1}{r^{d-2+\eta}} \qquad (r \rightarrow \infty),
\label{asymp2}
\end{equation}
where $(2-d) < \eta \le  2$ is a  ``critical'' exponent associated with $c(\bf r)$ for
hyperuniform systems that depends
on the space dimension. The Fourier transform of (\ref{asymp2})
yields
\begin{equation}
{\tilde c}({\bf k}) \sim -\frac{1}{k^{2-\eta}} \qquad (k \rightarrow 0),
\label{c-eta}
\end{equation}
which, when combined with (\ref{c}), yields the asymptotic form
of the structure factor
\begin{equation}
S({\bf k}) \sim k^{2-\eta} \qquad (k \rightarrow 0),
\label{S-eta}
\end{equation}
where $\eta=2-\alpha$ and $\alpha$ is the exponent defined
in relation (\ref{eq:Sk-scaling}).

 In what follows, it is assumed for concreteness, that the number density
is the control parameter. We define the following dimensionless density:
\begin{equation}
\phi=\rho v_1(D/2),
\label{phi}
\end{equation}
where $v_1(R)$ is given by (\ref{v1}) and $D$ is a characteristic ``microscopic" length scale.
The direct correlation in the vicinity of
a hyperuniform critical state with dimensionless density $\phi_c$, i.e., for $|\phi_c -\phi| \ll 1$, in sufficiently high dimensions
has  the following large-$r$ asymptotic form \cite{To03a,To18a}:
\begin{equation}
c({\bf r}) \sim \frac{\exp(-|{\bf r}|/\xi)}{|{\bf r}|^{d-2+\eta}},
\end{equation}
where $\xi$ is the {\it correlation length}. If the system approaches
a hyperuniform state from  below the critical density $\phi_c$, the correlation length
and inverse of the structure factor
at $k=0$, $S^{-1}(0)$, which is proportional to ${\tilde c}(0)$, are described by
the following scaling laws:
\begin{equation}
\xi  \sim  \left(1 - \frac{\phi}{\phi_c}\right)^{-\nu} \quad (\phi \rightarrow \phi_c^{-}),
\label{c-power}
\end{equation}
\begin{equation}
S^{-1}(0) \sim  \left(1-\frac{\phi}{\phi_c}\right)^{-\gamma}, \quad 
(\phi \rightarrow \phi_c^{-}), 
\label{S-power}
\end{equation}
where $\nu$ and $\gamma$ are nonnegative critical exponents. Observe
that the exponent $\gamma$ is a measure of how quickly a system approaches
a critical point. Combination of the three previous scaling laws
leads to the following interrelation between the exponents:
\begin{equation}
\gamma=(2-\eta)\nu.
\label{inter}
\end{equation}

The specific values of the critical exponents determine the {\it universality} class of
the hyperuniform system. It is noteworthy that all class III hyperuniform
systems are at critical points with the aforementioned scaling laws.
Specific examples of such a class III critical-point systems
are nonequilibrium absorbing-state models \cite{He15}.

\subsection{Growing Length Scale}

Another important length scale $\xi_c$ that can be extracted from the direct correlation
function and which grows as a hyperuniform state is approached
is defined by \cite{Ho12b}
\begin{equation}
\xi_c =[-{\tilde c}(k=0)]^{1/d}.
\label{xi-c}
\end{equation}
Thus, we see that $\xi_c$ is the $d$th root of the volume integral
of the direct correlation function $c({\bf r})$.
At a hyperuniform critical point, $\xi_c$ diverges to $+\infty$.
It was shown that a precursor to the hyperuniform maximally random jammed (MRJ) state of sphere packings \cite{To00b}
under compression was evident for densities far below the jamming density 
was reached, as reflected by this static growing length scale \cite{Ho12b}.
The quantity $\xi_c$ was also used to identify  length
scales in supercooled atomic liquid models  that substantially grow as the temperature decreased \cite{Ma13a}.

\subsection{Hyperuniformity Index}

The {\it hyperuniformity index} $H$ provides a measure of the nearness of a system to a hyperuniform
state, which is defined as
\begin{equation}
H \equiv \frac{S(0)}{S(k_{p})},
\end{equation}
where $k_{p}$ is the wavenumber $k$ associated with the largest peak height of the angularly averaged
structure factor. One may empirically deem a system to be nearly or effectively
hyperuniform if $H$ is roughly less than about $10^{-4}$ \cite{At16a}. The $H$ index 
has been profitably used to quantify the effective hyperuniformity of polymer systems \cite{Chr17,Ch18}, amorphous ices
\cite{Mar17}, states along the metastable extension of the hard-sphere systems away from jamming \cite{Zhou20}
and low-temperature states of ``quantizer" systems \cite{Kl19a,Ha20}.

\section{Hyperuniform and Nonhyperuniform Scaling Regimes}
\label{scaling}

For a large class of ordered and disordered systems, the number variance $\sigma^2(R)$ has the following large-$R$
asymptotic behavior~\cite{To03a, To18a}:
\begin{equation}
  \sigma^2(R)=2^d\phi\left[A\left(\frac{R}{D}\right)^d+B\left(\frac{R}{D}\right)^{d-1}+ o\left(\frac{R}{D}\right)^{d-1}\right] 
  \label{eq:sigma}
\end{equation}
where $\phi$ is the dimensionless density given by (\ref{v1}) and
$o\left(R/D\right)^{d-1}$ represents terms of lower order
than $\left(R/D\right)^{d-1}$. Moreover,
$A$ and $B$ are ``volume'' and ``surface-area''
coefficients, respectively, which can be expressed as the following volume integrals
involving the total correlation function $h({\bf r})$, respectively:
\begin{eqnarray}
  A&=&\lim_{|\mathbf{k}|\rightarrow 0}S(\mathbf{k})=1+ \rho \int_{\mathbb{R}^d} h({\bf r}) d{\bf r} \nonumber\\
&=& 1+  d\, 2^d\,\phi \int_0^\infty x^{d-1} h(x) dx ,
  \label{A}
\end{eqnarray}
\begin{eqnarray}
  B&=&-\frac{d\, \Gamma(d/2)\rho}{2\,\pi^{1/2} \,D\Gamma[(d+1)/2]} \int_{\mathbb{R}^d} |{\bf r}| h({\bf r}) d{\bf r} \nonumber\\
&=& -\frac{d^2\, 2^{d-1}\,\Gamma(d/2)\phi}{\pi^{1/2}\Gamma[(d+1)/2]} \int_0^\infty x^d\, h(x) dx, 
  \label{B}
\end{eqnarray}
and $x=r/D$ is a dimensionless distance. Here $h(r)$ is the radial function that depends on the distance $r \equiv |{\bf r}|$, which  results 
from averaging the vector-dependent quantity $h({\bf r})$, i.e.,
\begin{equation}
h(r) =\frac{1}{\Omega} \int_{\Omega} h({\bf r}) \, d\Omega,
\end{equation}
$d\Omega$ is the differential solid angle and 
\begin{equation}
\Omega = \frac{d \pi^{d/2}}{\Gamma(1+d/2)}
\end{equation}
is the total solid angle contained in a $d$-dimensional sphere.
In a perfectly hyperuniform system~\cite{To03a},
the non-negative volume coefficient vanishes, i.e., $A=0$, implying the sum rule
\begin{equation}
\rho \int_{\mathbb{R}^d} h({\bf r}) d{\bf r}=-1,
\end{equation}
such that the surface-area coefficient $B$ is nonnegative. Thus, such hyperuniform systems fall 
within class I, since the variance grows like the window surface area ($R^{d-1}$). 
On the other hand, when $A>0$ and $B=0$, the system is \textit{hyposurficial},
implying the sum rule
\begin{equation}
\int_0^\infty x^d h(x) dx=0.
\label{hypo-sum}
\end{equation}
  Examples of  hyposurficial systems include
ideal gases,  certain hard-core systems in $\mathbb{R}^d$ \cite{To03a},
 non-equilibrium phase transitions in amorphous ices \cite{Mar17}, and certain systems
with bounded pair interactions \cite{Zh20}. Appendix \ref{scalings} provides a more general
asymptotic expansion of the local number variance, which is used
to derive nonhyperuniform scaling laws. 

For a large class of nonhyperuniform disordered systems that are {\it sub-Poissonian} \cite{To21b}, such as 
fluids and colloids without particle clustering, the volume coefficient $A$ is often larger than the magnitude of the surface-area
coefficient $B$. For {\it super-Poissonian} configurations \cite{To21b}, such as the one described in Appendix \ref{super},
the magnitude of $B$ can be larger than $A$. 
When $A<1$ and $A<B$, $B$ is often positive, and the smallness of the ratio $A/B$ measures the degree
of hyperuniformity \cite{Mar17}. In a disordered system that is nearly hyperuniform, the 
inverse of this ratio, $B/A$ enables us to ascertain hyperuniform and nonhyperuniform
distance-scaling regimes of the variance $\sigma^2(R)$ as a function of $R$. 
The crossover value of $R_c$  between these two scaling regimes is determined 
by equating the first two terms of the large-$R$ asymptotic expansion of the local number variance, Eq. (\ref{eq:sigma}),
yielding the condition
\begin{equation}
R_c \approx \frac{B}{A}.
\end{equation}
For the range $R_0 < R < R_c$, where $R_0 \approx \rho^{-1/d}$ (roughly, equal to the mean-nearest neighbor distance),
the system exhibits hyperuniform scaling behavior, i.e.,
the variance is dominated by the surface-area scaling, $R^{d-1}$.
Clearly, the crossover value $R_c$ becomes infinite for perfectly hyperuniform systems
in which case $B/A=+\infty$. On the other hand, for $R>R_c$ and finite $B/A$, the system
begins to exhibit nonhyperuniform  scaling behavior, i.e.,
the variance is dominated by the surface-area scaling, $R^{d}$. The determination of  hyperuniform scaling regimes  could be especially useful in analyzing experimentally or computationally-generated systems that are necessarily of finite size $L$ such that $R < L/4$ \cite{To21b}.

\section{Fourier Representation of Surface-Area Coefficient}
\label{Fourier}

Here we derive a Fourier representation of the surface-area coefficient $B$
for any homogeneous many-particle system, whether hyperuniform or not, provided that the structure factor
meets certain mild conditions.  This representation will be especially useful when the scattering
intensity is available experimentally or if the structure factor
is analytically available, as it is when solving the Ornstein-Zernike integral equation (\ref{OZ}).

We define the
Fourier transform of a function $f({\bf r})$ that depends on the vector $\bf r$ in $d$-dimensional Euclidean space
$\mathbb{R}^d$ as follows:
\begin{eqnarray}
        \tilde{f}(\mathbf{k}) = \int_{\mathbb{R}^d} f(\mathbf{r}) \exp\left[-i(\mathbf{k}\cdot  \mathbf{r})\right] d\mathbf{r},
\end{eqnarray}
where $\bf k$ is a wave vector and  $(\mathbf{k} \cdot \mathbf{r}) = \sum_{i=1}^d k_i r_i$ is the conventional Euclidean inner product of two real-valued vectors.
The function $f({\bf r})$ can generally represent a tensor of arbitrary rank. When it is well-defined, the corresponding inverse Fourier transform is given by
\begin{eqnarray}
f(\mathbf{r}) = \left(\frac{1}{2\pi}\right)^d \int_{\mathbb{R}^d}       \tilde{f}(\mathbf{k}) \exp\left[i(\mathbf{k}\cdot  \mathbf{r})\right] d\mathbf{k}.
\label{Inverse}
\end{eqnarray}
If  $f$ is a radial function, i.e., $f$ depends only
on the modulus $r=|\mathbf{r}|$ of the vector $\bf r$,
its Fourier transform is given by
\begin{eqnarray}
{\tilde f}(k) =\left(2\pi\right)^{\frac{d}{2}}\int_{0}^{\infty}r^{d-1}f(r)
\frac{J_{\left(d/2\right)-1}\!\left(kr\right)}{\left(kr\right)^{\left(d/2\right
)-1}} \,d r,
\label{fourier}
\end{eqnarray}
where  $k=|{\bf k}|$ is the wavenumber or modulus of the wave vector $\bf k$
and $J_{\nu}(x)$ is the Bessel function of the first kind of order $\nu$.
The inverse transform of $\tilde{f}(k)$ is given by
\begin{eqnarray}
f(r) =\frac{1}{\left(2\pi\right)^{\frac{d}{2}}}\int_{0}^{\infty}k^{d-1}\tilde{f}(k)
\frac{J_{\left(d/2\right)-1}\!\left(kr\right)}{\left(kr\right)^{\left(d/2\right
)-1}} d k.
\label{inverse}
\end{eqnarray}

The Fourier representation of the local number variance for statistically homogeneous media,
which includes perfect crystals (under uniform translations of the crystals over their fundamental cells) \cite{To18a},
is given by
\begin{equation}
\sigma^2(R)=
2^d \phi R^d \Bigg[\frac{1}{(2\pi)^d} \int_{\mathbb{R}^d} S({\bf k}) 
{\tilde \alpha}({\bf k};{\bf R}) d{\bf k}\Bigg] ,
\label{var-1}
\end{equation}
where 
\begin{equation}
{\tilde \alpha}(k;R)= 2^d \pi^{d/2} \Gamma(1+d/2)\frac{[J_{d/2}(kR)]^2}{k^d}
\end{equation}
is the Fourier transform of the scaled intersection volume function $\alpha(r;R)$, which depends only on the magnitude
of the wavevector $k=|\bf k|$.    Using  the identity
\begin{equation}
\frac{1}{(2\pi)^d} \int_{\mathbb{R}^d} {\tilde \alpha}(k; R) d{\bf k}=1,
\label{identity}
\end{equation}
it follows from (\ref{var-1})  that 
\begin{equation}
\sigma^2(R)=  2^d \phi   S_0 \,R^d + 2^d \phi    \,R^d\Bigg[\frac{1}{(2\pi)^d} \int_{\mathbb{R}^d} [S({\bf k})-S_0] 
{\tilde \alpha}(k; R) d{\bf k}\Bigg] ,
\label{var-2}
\end{equation}
where 
\begin{equation}
S_0 \equiv \lim_{|{\bf k}| \to 0} S({\bf k}) =A=1+\rho \int_{\mathbb{R}^d} h({\bf r})\, d{\bf r},
\end{equation}
which implies
\begin{equation}
\rho {\tilde h}({\bf k=0})=S_0-1.
\end{equation}

Since ${\tilde \alpha}(k; R)$ is a radial function, depending only on the magnitude
of the wavevector, we can carry out the angular integration in the integral in (\ref{var-2}), yielding 
\begin{eqnarray}
\sigma^2(R) &=&  2^d \phi   S_0 \,R^d  \nonumber \\
\hspace{-0.2in}+&&\hspace{-0.2in}2^d \phi    \,R^d\Bigg[\frac{d \pi^{d/2}}{(2\pi)^d \Gamma(1+d/2)} \int_{\mathbb{R}^d} k^{d-1}[S(k)-S_0]
{\tilde \alpha}(k;R) dk\Bigg].
\label{var-3}
\end{eqnarray}
where the radial function $S(k)$ is given by
\begin{equation}
S(k) = \frac{1}{\Omega} \int_{\Omega} S({\bf k}) d\Omega.
\end{equation}
 For large $R$,
\begin{equation}
{\tilde \alpha}(k;R) \sim 2^{d+1} \pi^{d/2-1} \Gamma(1+d/2)\frac{ \cos^2[kR -(d+1)/4]}{R k^{d+1}}.
\label{asym}
\end{equation}
Combination of (\ref{var-3}) and (\ref{asym}) yields the following large-$R$ asymptotic
expansion:
\begin{eqnarray}
\sigma^2(R) &\sim&  2^d \phi   S_0 \,R^d  \nonumber \\
\hspace{-0.2in}+&&\hspace{-0.15in}2^d \phi    \,R^{d-1}\Bigg[\frac{2d}{\pi} \int_0^\infty \frac{S(k)-S_0}{k^2}  \cos^2[kR -\frac{d+1}{4}]\, dk \Bigg] \, \nonumber\\ &+&{\cal O}(R^{d-3}).
\label{var-4}
\end{eqnarray}
Using the identity
\begin{equation}
\lim_{L \to \infty} \frac{1}{L} \int_0^L \cos^2[kR -\frac{d+1}{4}]\, dR=\frac{1}{2}
\end{equation}
and  (\ref{var-4}), we obtain
\begin{equation}
\sigma^2(R) \sim  2^d \phi   S_0 \,R^d+  2^d \phi    \,R^{d-1}\Bigg[\frac{d}{\pi} \int_0^\infty \frac{S(k)-S_0}{k^2} \, dk\Bigg]  +{\cal O}(R^{d-3}).
\label{var-5}
\end{equation}

Comparing (\ref{var-5}) to (\ref{eq:sigma}) yields the desired Fourier representation of the
surface-area coefficient
\begin{equation}
B=\frac{d}{ \pi} \int_0^\infty \frac{S(k)-S_0}{k^2} \, dk.
\label{Fourier-B}
\end{equation}
Thus, this Fourier representation of the coefficient $B$ is bounded provided that the difference $[S(k)-S_0]$ tends to zero in the limit $k \to 0$ faster than linear in $k$.
Note that this condition will always be met by any structure factor that is analytic at the origin, since $[S(k)-S_0]$ must vanish
at least as fast as quadratically in $k$ as $k \to 0$.
Finally, we observe that if the system is hyposurficial ($B=0$), relation (\ref{Fourier-B})
leads to the integral condition
\begin{equation}
\int_0^\infty \frac{S(k)-S_0}{k^2} \, dk=0,
\end{equation}
which is the Fourier-space sum rule for hyposurficality that corresponds
to the direct-space sum rule (\ref{hypo-sum}).

\section{Toward Jammed States in Equilibrium Hard-Sphere Systems}
\label{free}

\begin{figure}[bthp]
\centerline{\includegraphics[width=3in,keepaspectratio,clip=]{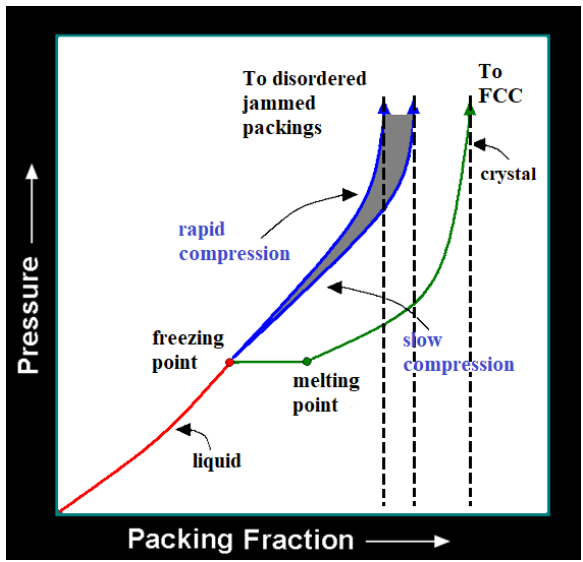}}
\caption{\footnotesize The isothermal phase behavior of the 3D hard-sphere model in the pressure-packing fraction plane, as adapted from
Ref. \onlinecite{To10c}. 
   An infinitesimal compression rate of the liquid traces out the thermodynamic equilibrium path (shown in red),
including a discontinuity resulting from the first-order freezing transition
to a crystal branch (shown in green) that ends at a maximally dense, infinite-pressure
jammed state.   Rapid compressions of the liquid (blue curves) generate a range
of amorphous metastable extensions of the liquid branch that jam
only at the their density maxima, which we show here must be perfectly hyperuniform states. }
\label{hsphere}
\end{figure}

We recall some well-known results for the equilibrium phase behavior of identical hard spheres
of diameter $D$. The pressure of an equilibrium hard-sphere system in any space dimension
can be expressed in terms of the contact values of the direct correlation function
from the right and left sides  via the Ornstein--Zernike equation \cite{To02a}:
\begin{equation}
\frac{p}{\rho k_B T}=1+2^{d-1}\phi \,[c(D^{+})-c(D^{-})].
\label{Z2}
\end{equation}
Here the dimensionless density $\phi$ is to be interpreted as the {\it packing fraction},
i.e., the fraction of space covered by the spheres.
Figure \ref{hsphere} schematically shows the three-dimensional (3D) phase behavior
in the $\phi$-$p$ plane. Three different isothermal  densification paths by which a hard-sphere liquid
may jam are shown. At sufficiently low densities,  an infinitesimally slow
compression of the system at constant temperature defines a  thermodynamically stable liquid branch for packing
fractions up to the ``freezing" point ($\phi \approx 0.494$). Increasing the density
beyond the freezing point putatively results in an entropy-driven first-order phase transition \cite{Al57,Frenk99}
to a crystal branch that begins at the melting point ($\phi \approx 0.55$).
 Slow compression of the system along the crystal branch ends at the
jammed state corresponding to the fcc lattice packing \cite{Mau99} with  $\phi = \pi/\sqrt{18}=0.74048 \ldots$.
Rapid compressions of the liquid while suppressing
some degree of local order  can avoid
crystal nucleation (on short time scales) and produce a range
of amorphous metastable extensions of the liquid branch that jam
only at the their density maxima. The faster the compression rate, the lower is the jammed
density.  Presumably, the metastable branch produced by the
most rapid compression rate with a terminal density consistent with strict jamming corresponds to
the maximally random jammed (MRJ) state with $\phi \approx 0.64$ \cite{To00b,To10c}.
The MRJ state under the strict-jamming constraint \cite{To01b,Do04c} is a prototypical glass
\cite{To07} in that it is maximally disordered (according to a variety of order metrics) without any  long-range order (Bragg peaks)
and perfectly rigid, i.e., the elastic moduli are  unbounded \cite{To03c,To10c}.

In the immediate vicinity of a jammed state in $\mathbb{R}^d$ with packing fraction $\phi_J$, 
the set of particle displacements  that  are  accessible  to  the  
packing  approaches  a convex-limiting  polytope (because the impenetrability constraints become linear)  and
free-volume theory \cite{Sa62,St69,Do05c} predicts that the pressure $p$ has the following exact asymptotic form:
\begin{equation}
\frac{p}{\rho k_B T} \sim \frac{d}{1-\phi/\phi_J} \qquad (\phi \to \phi_J).
\label{free-volume}
\end{equation}
So far as the limiting polytope picture is concerned, the extremely narrow connecting filaments that in principle connect the jamming neighborhoods have so little measure that they do not overturn the free-volume leading behavior of the pressure, even in the infinite-system limit \cite{To10c}.  
Although there is no rigorous proof yet for this claim, all numerical evidence strongly suggests that it is correct
as the jamming state is approached either along  the crystal \cite{Sa62,St69} or metastable branch \cite{Do05c}. Assuming that the system dynamics remain ergodic in the vicinity
of a jammed state, we can apply the fluctuation-compressibility relation (\ref{comp}) 
along with the free-volume form (\ref{free-volume}) to yield the corresponding
asymptotic relation for $S^{-1}(0)$:
\begin{equation}
S^{-1}(0) \sim \frac{d}{(1-\phi/\phi_J)^{2}} \qquad (\phi \to \phi_J).
\label{S-inverse}
\end{equation}
Thus, we see that in the limit $\phi \to \phi_J$, the system becomes perfectly hyperuniform,
which appears to validate a more general conjecture of Torquato and Stillinger \cite{To03a},
as elaborated below.

It is important to note that ergodicity condition required to obtain (\ref{S-inverse})
implies that the system must be defect-free. To stress this point further, consider
an fcc packing in $\mathbb{R}^3$ at the jammed state with packing  fraction
$\phi = \pi/\sqrt{18}=0.74048 \ldots$. Next, shrink each sphere by an
infinitesimal uniform amount (such that the impenetrability constraints are linerizable) and randomly remove
a small but statistically significant fraction of spheres so that the resulting vacancies
are well-separated from one another. This vacancy-riddled packing  remains in its jamming basin. Now, slowly compress it 
until it jams, whereby the system pressure diverges. It is known that
the structure factor $S({\bf k})$ at the origin of a nearly hyperuniform system with vacancies
is proportional to the concentration of vacancies \cite{Ki18a} and hence cannot be hyperuniform,
as predicted by (\ref{S-inverse}). Thus, the free-volume form for the pressure (\ref{free-volume}) cannot
apply to this nonequilibrium vacancy-riddled but strictly jammed state.

More generally, Torquato and Stillinger \cite{To03a} suggested that certain defect-free strictly jammed packings of identical spheres
are hyperuniform. Specifically, they conjectured that any strictly jammed saturated infinite packing of identical spheres is hyperuniform. A saturated
packing of hard spheres is one in which there is no space available to add another sphere. What is the rationale for
such a conjecture? First, it recognizes that mechanical rigidity is a necessary but not sufficient condition for hyperuniformity.
Indeed, requiring the saturation property in the conjecture eliminates the class of strictly jammed crystal states
that are riddled with vacancies, as per the aforementioned example. Moreover, the  conjecture excludes packings that may
have a rigid backbone but possess ``rattlers” (particles that are not locally jammed but are free to move about a confining
cage) because a strictly jammed packing, by definition,  cannot contain rattlers \cite{Do05c,To10c}. Typical packing protocols that have generated disordered jammed packings tend to contain a small concentration of rattlers \cite{To10c,Pa10},
and hence the entire (saturated) packing cannot be deemed to be jammed. Therefore, the conjecture cannot apply to current
numerically-generated disordered packings, even if the structure factor at the origin is very small, e.g., an $H$ index of the order
of $10^{-4}$ for 3D MRJ-like sphere packing \cite{Do05d}. Thus, rattler-free disordered jammed sphere packings
are expected to be perfectly hyperuniform. Indeed, if the free-volume theory applies along the metastable extension 
ending at the MRJ state, rattlers cannot be present (due to a type of constrained ergodicity on time scales
much less than relaxation times) and hence
should be perfectly hyperuniform. It has been suggested that the ideal MRJ state is rattler-free, implying
that the packing is more disordered without the presence of rattlers \cite{At16a}.

The consequences of the free-volume theory
requires a modification of the Torquato-Stillinger conjecture because the former 
eliminates defects of {\it any} type in the jamming limit and the saturation
condition may not prohibit all defect types. For example, the saturation property may not exclude dislocations
in strictly jammed crystal states. A more refined variant of the conjecture is the following
statement: Any strictly jammed infinite packing of identical spheres that is defect-free is hyperuniform.

In Appendix \ref{sum-rule}, we obtain exact   sum rules and the   exact large-$k$ asymptotic behaviors of the structure
factors of certain general packings of identical spheres in $\mathbb{R}^d$, whether in equilibrium or not.

\section{Equilibrium Hard Rods}
\label{rods}

We first structurally characterize one-dimensional systems of identical hard rods of length $D$
in equilibrium as a function of the packing fraction $\phi$ up to the jammed, hyperuniform
state ($\phi=1$).  This hyperuniform state is the integer lattice $\mathbb{Z}$
and hence cannot be regarded to be a critical hyperuniform state, as described
in Sec. \ref{inverted}.

Percus \cite{Pe64} obtained the following exact expression for the direct correlation
function for an equilibrium hard-rod system at packing fraction $\phi$:
\begin{equation}
c(r)=-\Theta(D-r) \frac{1-\phi r/D}{(1-\phi)^2} 
\end{equation}
where $\Theta(x)$ is the Heaviside step function.
Combination of this relation with (\ref{Z2}) yields the well-known
expression for the pressure of the hard-rod system \cite{To36}:
\begin{equation}
\frac{p}{\rho k_B T}=\frac{1}{1-\phi},
\end{equation}
which we see takes the free-volume form (\ref{free-volume}) for the entire range
of packing fractions, not just near the jammed state $\phi_J=1$, which
must be hyperuniform.  The Fourier transform of $c(r)$ is given by
\begin{equation}
{\tilde c}(k)=  \frac{2\left[ \phi\, (\cos(kD)-1) +kD\,\sin(kD)(\phi-1)\right]}{(1-\phi)^2 (kD)^2}.
\label{c-rods}
\end{equation}
Using (\ref{c}), the corresponding structure factor is given by
\begin{equation}
S(k)=\frac{1}{1-\frac{\displaystyle 2\,\phi\left[ \phi\, (\cos(kD)-1) +kD\,\sin(kD)(\phi-1)\right]}{\displaystyle (1-\phi)^2 (kD)^2}}.
\label{S-rods-eq}
\end{equation}
It is seen that the asymptotic large-$k$ behavior of $S(k)$ is consistent with the general
expression  (\ref{S-largek-1}) for $d=1$, where the contact value $g_2(D^+)=(1-\phi)^{-1}$
is bounded for all $\phi$, except at its maximal value $\phi=1$.

\begin{figure}[H]
\centering{\includegraphics[width = 0.5\textwidth,clip=]{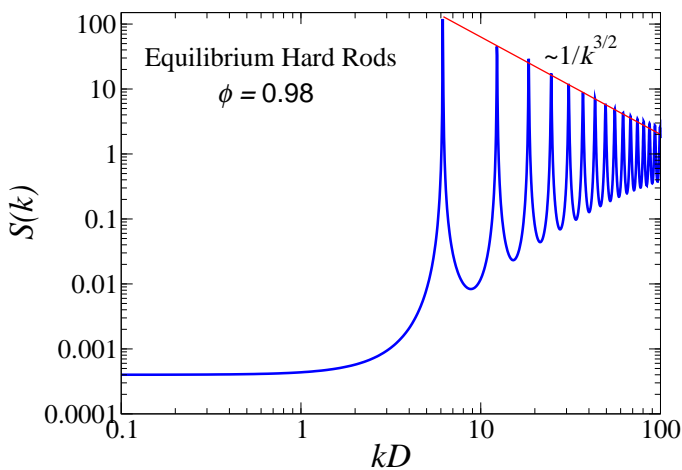}}
\caption{The structure factor $S(k)$ versus the dimensionless wavenumber $kD$ for a equilibrium
hard-rod system near the jammed state ($\phi=0.98$). The envelope of the largest peak values decays with wavenumber
approximately as the power law $1/k^{3/2}$.  }
\label{S-rods}
\end{figure}

The structure factor $S(k)$ is plotted in  Fig. \ref{S-rods} for the
hard-rod system near the jammed state with $\phi=0.98$. At such high densities,
its peak values occur at wavenumbers that are approximately integer multiples
of $2 \pi/D$, which is exactly the case for the integer lattice (jammed state) with lattice constant $D$. The envelope of the largest peak values decays with wavenumber
approximately as the power law $1/k^{3/2}$. It immediately follows from (\ref{S-rods-eq}) that the volume coefficient
$A$, defined by (\ref{A}), is exactly given by
\begin{equation}
A=S(0)=(1-\phi)^2
\label{A-rods}
\end{equation}
The  surface-area coefficient $B$ is easily obtained for all $\phi$
from its Fourier representation  (\ref{Fourier-B})  and use of the exact structure factor formula (\ref{S-rods-eq})
by high-precision numerical integration. These results for $B$ are in excellent agreement with previous results
using the direct-space representation of $B$ \cite{To03a}. A simple and highly accurate formula for $B$ 
is given by
\begin{equation}
B=\frac{\phi}{12}(6-8\phi+3\phi^2).
\end{equation}
This formula is obtained by using exact results at low and high packing fractions and assuming that $B$ is a cubic polynomial in $\phi$ (without a constant term).
Specifically, the linear and quadratic terms are determined by the exact expansion of  $S(k)-S_0$ in powers of $\phi$ through second order in $\phi$, as
determined from the the exact formula (\ref{S-rods-eq}). The remaining cubic term is found by using the fact that $B$ is exactly equal to $1/12$ for the integer 
lattice at $\phi=1$ \cite{To03a}. Thus,  combination of the aforementioned formulas for $A$ and $B$, yields
the following excellent approximation for the ratio
\begin{equation}
\frac{B}{A}=\frac{\phi(6-8\phi+3\phi^2)}{12(1-\phi)^2}.
\label{B/A-rods}
\end{equation}

\begin{figure}[H]
\centering{\includegraphics[width = 0.5\textwidth,clip=]{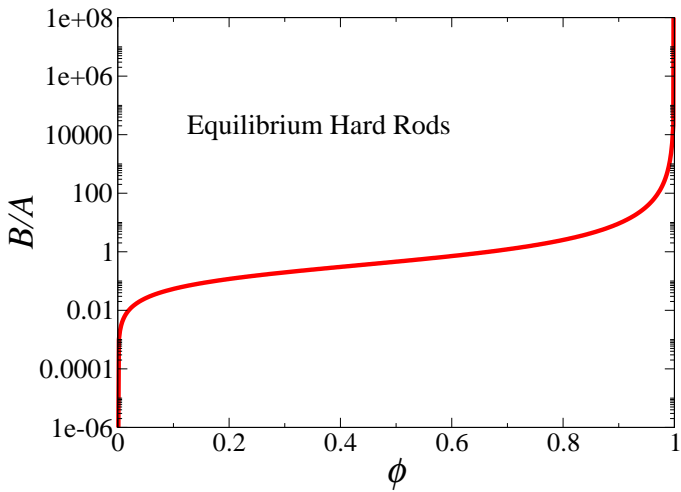}}
\caption{The ratio of the surface-area coefficient to volume coefficient, $B/A$, as
a function of the packing fraction $\phi$ for  equilibrium
hard-rod systems.}
\label{A-B-rods}
\end{figure}

\begin{figure}[H]
\centering{\includegraphics[width = 0.5\textwidth,clip=]{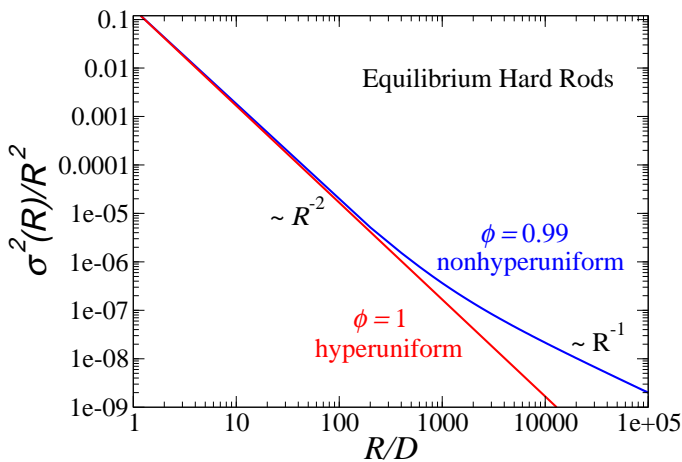}}
\caption{The scaled variance $\sigma^2(R)/R^2$ versus $R/D$, as computed from relation (\ref{eq:sigma}), for the jammed hyperuniform
state with $\phi=1$ (as found in Ref. \cite{To03a}) and for a nonhyperuniform state near jamming with $\phi=0.99$, where $R_c=B/A =841.7474996$.
Because the variance is bounded for the hyperuniform state \cite{To03a}, $\sigma^2(R)/R^2$ decreases with $R$ for large $R$ like $1/R^2$. }
\label{var-rods}
\end{figure}

The ratio $B/A$ is plotted in  Fig. \ref{A-B-rods}
as a function of $\phi$. It is seen that $B$ is about ten times larger than $A$ at $\phi=0.9$
and the ratio dramatically diverges to infinity as $\phi$  tends to unity. For example, at $\phi=0.99$,
$B$ is about 842 times larger than $A$. Referring to Sec. \ref{scalings}, the magnitude of the ratio $B/A$ determines the hyperuniformity 
scaling regime for the local number variance $\sigma^2(R)$ to be those window radii in the range 
${\cal O}(\rho^{-1/d}) < R < R_c = B/A$. For $R$ beyond ${\cal O}(B/A)$, the variance should display nonhyperuniform
scaling of the number variance.  These scaling regimes are verified in the plot of the variance
at $\phi=0.99$ shown in Fig. \ref{var-rods}. The actual variance, as determined from  the exact relations (\ref{var-1}) 
and (\ref{S-rods-eq}), at such a high packing fraction is characterized by small-scale fluctuations around
some average values, and because such variations visually obscure the scaling regimes,
we considered employing the cumulative running average, defined in Ref. \cite{Ki17},  to smooth out the small-scale
oscillations around the global values. Because the cumulative average plot on the scales shown in Fig. \ref{var-rods}
is very similar to one obtained from the asymptotic expansion (\ref{eq:sigma}), we employ the later, for simplicity, to 
show the scaling regimes.


\begin{figure}[H]
\centering{\includegraphics[width = 0.5\textwidth,clip=]{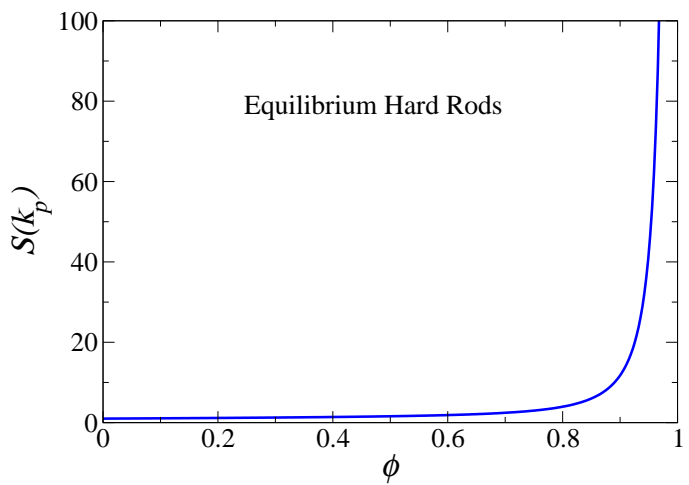}}
\caption{The largest peak value of the structure factor $S(k_{p})$ versus the packing fraction $\phi$ for  equilibrium
hard-rod systems. }
\label{S-peak}
\end{figure}

\begin{figure}[H]
\centering{\includegraphics[width = 0.5\textwidth,clip=]{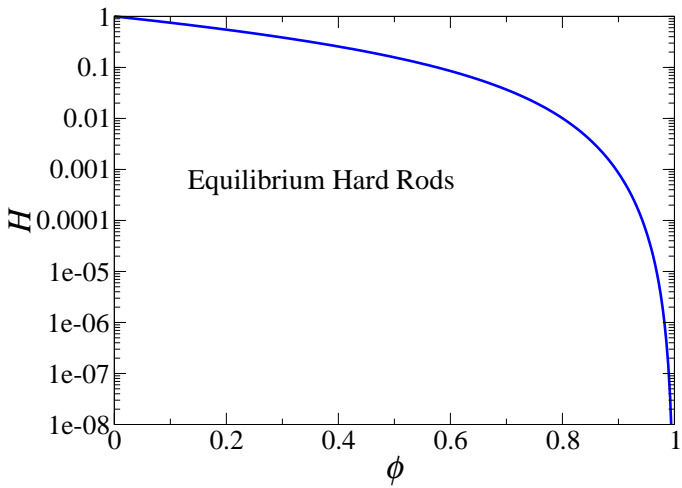}}
\caption{The hyperuniform index $H$ versus the packing fraction $\phi$ for  equilibrium
hard-rod systems. The rapid growth of $\xi_c$ with increasing packing fraction is readily
apparent.}
\label{H}
\end{figure}

It easily follows from formula (\ref{S-rods-eq}) that the largest peak value of the structure factor $S(k_p)$
through first-order in the packing fraction is exactly  given by
\begin{eqnarray}
S(k_p)&=&1-\frac{\sin(k_p D)}{k_p D} \phi  + {\cal O}(\phi^2)\nonumber \\
&=& 1 + (0.434464192\ldots) \phi  + {\cal O}(\phi^2),
\label{S-k}
\end{eqnarray}
where 
\begin{equation}
k_p D = \frac{3}{4}\pi+\frac{1}{4}\sqrt{9\pi^2-16} = 4.489654702\ldots \,.
\end{equation}
It is highly nontrivial to derive an analytical expression for $S(k_p)$ for arbitrary $\phi$.
Of course, we know that $S(k_p)$ must increase as $\phi$ increases
and diverges in the limit that the jamming density is approached from below, i.e., $\phi \to 1^{-}$, which should be
distinguished from the case when $\phi=1$ in which $S(k)$ consists of Dirac
delta functions located at wavenumbers that are multiples of $2\pi/D$. 
The first peak value of the structure factor, which is also the largest value, at a fixed value
of the packing fraction is plotted in Fig. \ref{S-peak}
as a function of  $\phi$, which of course must diverge
at the hyperuniform jammed state $\phi=1$. The following rational function provides
an excellent fit to the data for all $\phi$ up to the jamming density:
\begin{equation}
S(k_{p})=\frac{1+a_1\phi+a_2\phi^2+a_3\phi^3}{(1-\phi)^2},
\end{equation}
where $a_1=-1.565536$, $a_2=0.470399$ and $a_3=0.196196$.
This fit uses the exact small-$\phi$ expansion (\ref{S-k}) to ascertain the coefficient $a_1$. 
This accurate approximation  when combined with exact formula (\ref{A-rods}) for $S(0)$
yields the following expression for the  hyperuniformity index: 
\begin{equation}
H= \frac{(1-\phi)^4}{1+a_1\phi+a_2\phi^2+a_3\phi^3},
\label{H-rods}
\end{equation}
 which is plotted
in Fig. \ref{H} as function of $\phi$.

\begin{figure}[H]
\centering{\includegraphics[width = 0.5\textwidth,clip=]{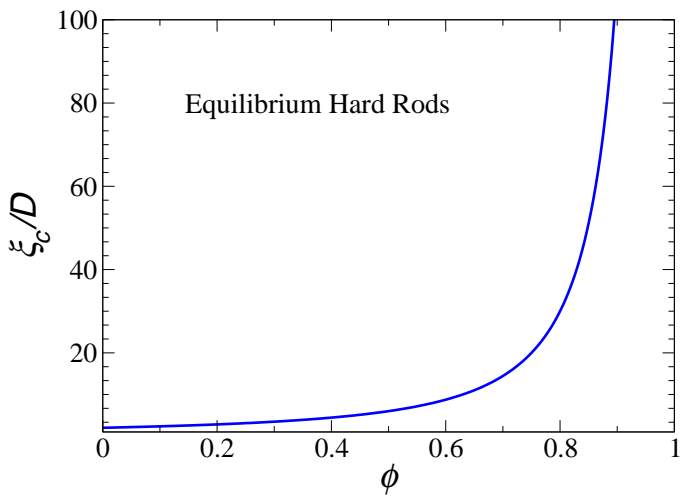}}
\caption{The dimensionless length scale $\xi/D$ versus the packing fraction $\phi$ for  equilibrium
hard-rod systems.}
\label{length-scale}
\end{figure}

Using  (\ref{c-rods}), it immediately follows that the length scale $\xi_c$, defined by (\ref{xi-c}), 
for equilibrium hard rods is given by
\begin{equation}
\xi_c=-{\tilde c}(k=0)=\frac{(2-\phi)D}{(1-\phi)^2}.
\label{xi-rods}
\end{equation}
Figure \ref{length-scale} clearly shows that this length scale grows appreciably with increasing packing fraction. It is already
about an order of magnitude greater than the length of a rod $D$ at $\phi=0.6$
and rapidly takes on even larger values as $\phi$ increases further,
eventually diverging to infinity in the limit $\phi\to 1$.

It is useful to compare and contrast the growth rate of the three descriptors $B/A$,  $\xi_c$
and $H^{-1}$ as the packing fraction $\phi$ increases and approaches the jammed hyperuniform state. We see that the numerators 
of the formulas  (\ref{B/A-rods}) and (\ref{xi-rods}) for moderate to high values of $\phi$ are of order unity, 
and hence  $B/A$ and $\xi_c$ have the same scaling behavior as the hyperuniform state of the hard-rod system is approached, namely,
\begin{equation}
\frac{B}{A} \sim \frac{\xi_c}{D} \sim \frac{1}{(1-\phi)^2}.
\end{equation}
This is to be contrasted with the inverse of the hyperuniformity index $H^{-1}$, 
which grows exponentially faster, i.e., we see from  (\ref{H}) that it grows
like the square of the other two quantities:
\begin{equation}
 H^{-1} \sim \left(\frac{B}{A}\right)^2 \sim \left(\frac{\xi_c}{D}\right)^2  \sim \frac{1}{(1-\phi)^4}.
\end{equation}


\section{Sticky Hard Spheres}
\label{stick}

We utilize a particular {\it $g_2$-invariant process} to analytically
study the approach of another disordered system to a hyperuniform  but unjammed state as a function of the relevant control parameter,
the number density.  A $g_2$-invariant process is one in which a chosen nonnegative form for
the pair correlation function $g_2$ remains invariant over a nonvanishing density range while keeping
all other relevant macroscopic variables fixed \cite{To02c,To03a}. The upper
limiting ``terminal'' density is the point above which
the nonnegativity condition on the structure factor, i.e., $S({\bf k}) \ge 0$ for all $\bf k$, 
would be violated. Thus, at the terminal or critical density, the system is hyperuniform
if the minimum in $S({\bf k})$ occurs at the origin and it is realizable. 
This optimization problem has deep connections to the sphere-packing
problem. Specifically, it is the linear program (LP) lower bound on the maximal packing 
density that is dual to the Elkies-Cohn LP upper bound formulation \cite{Co03}. A certain $g_2$-invariant
test function was employed to conjecture that the densest sphere packings in very high
space dimensions are disordered, rather than ordered as in low dimensions \cite{To06b}.

\begin{figure}[H]
\centering{\includegraphics[width = 0.4\textwidth,clip=]{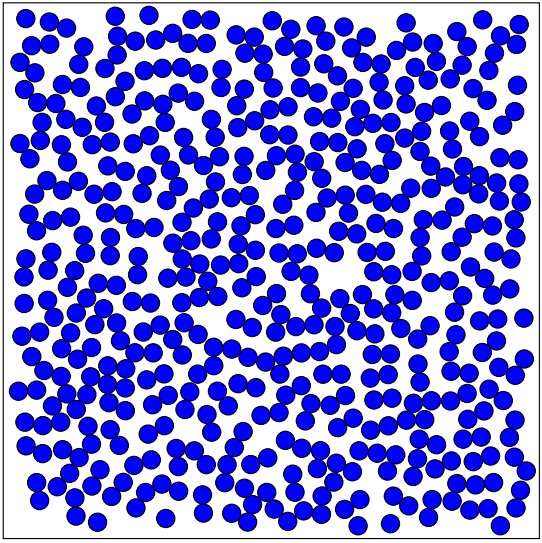}}
\caption{A numerically generated two-dimensional configuration of 500 ``sticky" hard disks that realizes
the step plus delta function pair correlation function at the terminal packing fraction $\phi_c=1/2$, as adapted from Ref. \cite{Uc06a}. 
This critical hyperuniform state  consists only of dimers, i.e., $Z=1$, and hence is unjammed, in contrast to hyperuniform states
in the phase diagram of hard spheres, as shown in Fig. \ref{hsphere}.}
\label{sticky}
\end{figure}

Here we specifically consider the $g_2$-invariant process defined by
a pair correlation function  that imposes a hard-core
constraint via a unit step function plus a delta function contribution
that acts at contact $r=D$:
\begin{equation}
g_2(r)=\Theta(r-D)+ \frac{Z}{\rho s_1(D)}\delta(r-D), 
\label{step-delta}
\end{equation}
where $Z$ is the nonnegative  average contact coordination number and 
\begin{equation}
s_1(r)=\frac{d \pi^{d/2} r^{d-1}}{\Gamma(1+d/2)}
\end{equation}
is the surface area of a sphere of radius $r$.  It was previously shown that
this ``sticky-sphere" pair correlation function is numerically realizable, to an excellent approximation,
up to the hyperuniform terminal packing fraction for $d=2$ (see Fig. \ref{sticky}), implying that
such pair correlation functions are also realizable in all higher dimensions, as dictated
by the {\it decorrelation principle} \cite{To06b}.

Here we collect previously established key results for this $g_2$-invariant process \cite{To02c,To03a} in order
to characterize how such systems for any $d$ approach a hyperuniform state
as the packing fraction increases up to the critical terminal value $\phi_c$. It was shown that
\begin{equation}
Z=\frac{2^d d}{d+2}\phi
\label{Z}
\end{equation}
is obeyed in order to constrain the location of the minimum
of the structure factor to be at ${\bf k=0}$, where the packing $\phi$ lies
in the range $0 \le \phi \le \phi_c$, and
\begin{equation}
\phi_c=\frac{d+2}{2^{d+1}}
\label{phic-sticky}
\end{equation}
is the terminal or critical packing fraction, which has the same asymptotic form
as a lower bound on the maximal packing density for lattice packings derived by Ball \cite{Ball92}.
At the hyperuniform critical point, the contact number $Z_c=d/2$, indicating that
such sticky-sphere systems are never jammed in any dimension.

\begin{figure}[H]
\centering{\includegraphics[width = 0.5\textwidth,clip=]{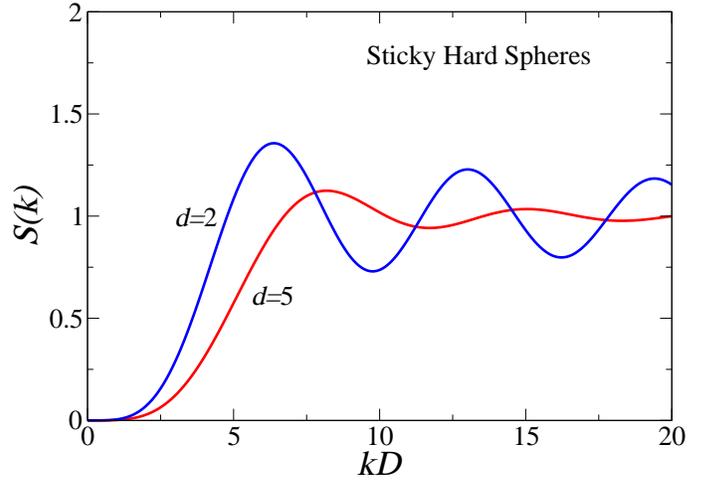}}
\caption{The structure factor $S(k)$ versus the dimensionless wavenumber $kD$ for sticky hard spheres
in two and five dimensions at their respective hyperuniform states, i.e., $\phi=\phi_c$, where
$\phi_c$ is given by (\ref{phic-sticky}). }
\label{S-sticky-fig}
\end{figure}

The corresponding structure factor for any $d$
and  $\phi$ in the range $0 \le \phi \le \phi_c$ is given by
\begin{equation}
S(k)=1+
\frac{2^{d/2} \Gamma(2+d/2) }{(kD)^{(d/2)-1}}
\left(\frac{\phi}{\phi_c}\right) \Bigg[\frac{J_{(d/2)-1}(kD)}{d+2}-\frac{J_{d/2}(kD)}{kD}\Bigg].
\label{S-sticky}
\end{equation}
An important mathematical property of the structure factor for such sticky hard spheres
is that its extrema for a fixed $d$ are independent of the packing
fraction. For example, for $d=2$ and $d=5$, $k_{p}D=6.38015\ldots$ and $k_{p}D=8.18255\ldots$, respectively.
It is seen that the asymptotic large-$k$ behavior of $S(k)$ is consistent with the general
expression  (\ref{S-largek}) for any $d$, where $Z$ is given by (\ref{Z}).
Relation (\ref{S-sticky}) leads to the power law
\begin{equation}
S^{-1}(0)=\left(1-\frac{\phi}{\phi_c} \right)^{-1}, \qquad \phi
\rightarrow \phi_c^{-}, 
\label{S-sticky-inverse}
\end{equation}
which upon comparison to (\ref{S-power})  yields the critical exponent $\gamma=1$. 
Figure \ref{S-sticky-fig} shows the structure factor $S(k)$ versus $kD$ for sticky hard spheres
for $d=2$ and $d=5$ at their respective hyperuniform states. We see that structure factor reflects
strong decorrelation in going from two to five dimensions, which
is consistent with the decorrelation principle \cite{To06b}.

\begin{figure}[H]
\centering{\includegraphics[width = 0.5\textwidth,clip=]{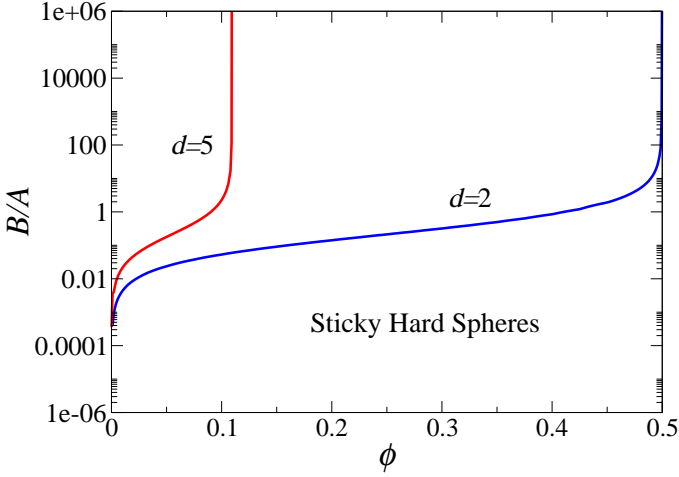}}
\caption{The ratio of the surface-area coefficient to volume coefficient, $B/A$, as
a function of the packing fraction $\phi$ for sticky hard spheres
in two and five dimensions.  }
\label{B-A-sticky}
\end{figure}

In any space dimension $d$, the volume and surface-area coefficients for $0 \le \phi \le \phi_c$ are 
respectively  given by  
\begin{equation}
A=S(k=0)=1-\frac{\phi}{\phi_c}
\label{A-sticky}
\end{equation}
and
\begin{equation}
B=\frac{ d^2 \Gamma(d/2)}
{8\,\sqrt{\pi}\,\Gamma((d+3)/2)}\frac{\phi}{\phi_c}.
\end{equation}
Thus, ratio $B/A$ is given by
\begin{equation}
\frac{B}{A}= \frac{ d^2 \Gamma(d/2)}
{8\,\sqrt{\pi}\,\Gamma((d+3)/2)}\frac{\phi/\phi_c}{(1-\phi/\phi_c)}.
\label{B-A-stick}
\end{equation}
We see that apart from $d$-dimensional constants, the behavior
of the ratio $B/A$ is exactly the same across dimensions
when the packing fraction $\phi$ is scaled by the terminal packing
fraction $\phi_c$. The ratio $B/A$ versus $\phi$ is
plotted in Fig. \ref{B-A-sticky}
for $d=2$ and $d=5$ for all $\phi$ up to their respective critical points.

\begin{figure}[H]
\centering{\includegraphics[width = 0.5\textwidth,clip=]{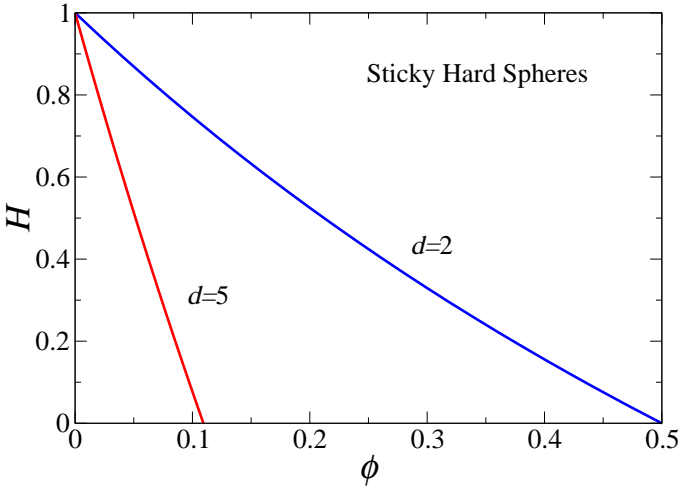}}
\caption{The hyperuniformity index $H$  as
a function of the packing fraction $\phi$ for sticky hard spheres
in two and five dimensions.}
\label{H-sticky-fig}
\end{figure}

We see from Fig. \ref{S-sticky-fig} that the peak value of the structure factor
is not strongly dependent on the space dimension and is of order unity. Thus, the
hyperuniformity index $H$ is primarily determined by the inverse of $A$ [cf. \ref{A-sticky})] and so
\begin{equation}
H \sim \frac{1}{1-\phi/\phi_c}.
\label{H-sticky}
\end{equation}
The actual values of $H$ are plotted in Fig. \ref{H-sticky-fig} as a function of $\phi$
for $d=2$ and $d=5$ up to their respective critical points.

Combination of (\ref{c}) and (\ref{S-sticky}) yields the following expression
for the the Fourier transform of the direct correlation for sticky spheres:
\begin{equation}
{\tilde c}(k)= \frac{\displaystyle \frac{(2\pi)^{d/2} D^d }{(kD)^{(d/2)-1}}
 \Bigg[\frac{\displaystyle J_{(d/2)-1}(kD)}{d+2}-
\frac{\displaystyle J_{d/2}(kD)}{kD}\Bigg]}
{\displaystyle 1+
\frac{2^{d/2} \Gamma(2+d/2) }{(kD)^{(d/2)-1}}
\left(\frac{\phi}{\phi_c}\right) \Bigg[\frac{J_{(d/2)-1}(kD)}{d+2}-
\frac{J_{d/2}(kD)}{kD}\Bigg]}
\end{equation}
It immediately follows that at zero wavenumber, we have
\begin{equation}
{\tilde c}(0)=-2v_1(D) 
 \left(1-\frac{\phi}{\phi_c} \right)^{-1},
\label{c-step-delta}
\end{equation}
and, hence, the length scale defined by (\ref{xi-c}) yields 
\begin{equation}
\xi_c = [2 v_1(D)]^{1/d}  \left(1-\frac{\phi}{\phi_c} \right)^{-1/d},
\label{xi_c-sticky}
\end{equation}
indicating that it grows more slowly as the space dimension increases.
Figure \ref{length-scale} shows how the length scale $\xi_c$ grows
with $\phi$ for $d=2$ and $d=3$.

The length scale $\xi_c$ should be contrasted with the correlation length
$\xi$, defined by (\ref{c-power}), which,  when combined with relation (\ref{c-step-delta}), yields
\begin{eqnarray}
\xi &=& \frac{D}{[ 8(d+2)(d+4)\phi_c]^{1/4}} \left(1-\frac{\phi}{\phi_c} \right)^{-1/4}, 
\qquad \phi \rightarrow \phi_c^{-}.
\label{xi-step-delta}
\end{eqnarray}
Comparison of (\ref{xi-step-delta}) to
the power law (\ref{c-power}) yields the exponent $\nu =1/4$.
It should not go unnoticed that since the exponents $\gamma$, $\eta$ and $\nu$ 
are either integers or a simple fraction, independent of dimension, these
systems always behave in a mean-field manner at their critical points.

\begin{figure}[H]
\centering{\includegraphics[width = 0.5\textwidth,clip=]{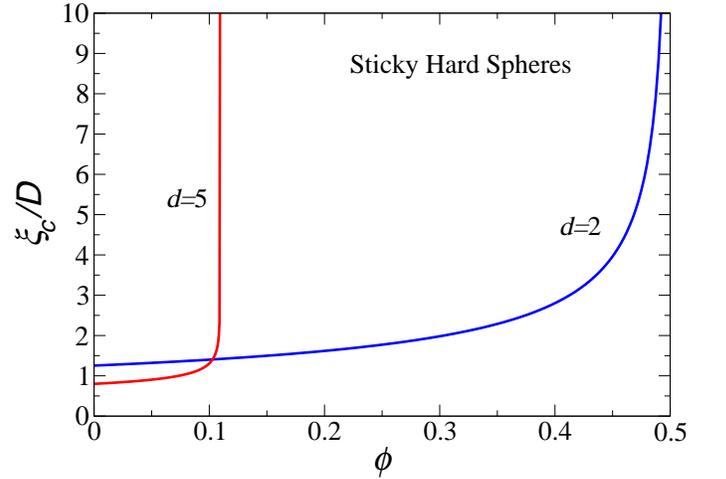}}
\caption{The dimensionless length scale $\xi_c/D$ versus the packing fraction $\phi$ for sticky hard spheres
in two and five dimensions.}
\end{figure}

All three descriptors $B/A$, $H^{-1}$ and $\xi_c$ are increasing functions of
the packing fraction $\phi$, but how do they correlate with one another? From formulas  (\ref{B-A-stick}) and (\ref{H-sticky}),
we see $B/A$ and $H^{-1}$ have the same scaling behavior as the critical hyperuniform state 
of the sticky hard-system is approached, namely,
\begin{equation}
\frac{B}{A} \sim H^{-1} \sim \frac{1}{(1-\phi/\phi_c)}.
\end{equation}
According to (\ref{xi_c-sticky}), the length scale $\xi_c$ for $d \ge 2$ grows more slowly than the other
two quantities as the critical state is approached, i.e., 
\begin{equation}
 (\xi_c)^d \sim \frac{B}{A} \sim H^{-1}.
\end{equation}

\section{Stealthy Disordered Ground States}
\label{stealthy}

We are also interested in theoretically understanding how vibrational modes in excited low-temperatures
states of matter impact the approach to ground states, which are necessarily hyperuniform (see Sec. \ref{f-c}),
as the temperature $T$ tends to zero. For this purpose, we consider
stealthy hyperuniform many-particle systems \cite{Uc04b,Ba08}, which are a subclass of class I hyperuniform systems \cite{To15} in which 
the structure factor is  zero for a range of wavevectors around the origin, i.e.,
\begin{equation}
S({\bf k}) =0  \qquad 0 \le |{\bf k}| \le K,
\end{equation}
where $K$ is some positive constant. Such systems are called
``stealthy"  because they completely suppress single scattering of incident radiation for
the wave vectors within an exclusion sphere of radius $K$ centered at the origin
in Fourier space. It has been shown that certain bounded (soft) long-ranged pair interactions
have classical ground states that are stealthy and hyperuniform \cite{To15}.
The nature of the ground-state configuration manifold (e.g., the degree of order) associated with such stealthy
pair potentials depends on the number of constrained
wave vectors. The dimensionless parameter $\chi$ measures the relative fraction of constrained
degrees of freedom compared to the total number of degrees of freedom and, counterintuitively, is inversely proportional to
the number density \cite{To15}; specifically, it is explicitly given by the formula
\begin{equation}
\rho \,\chi =\frac{v_1(K)}{2d\,(2\pi)^d},
\label{rho-chi}
\end{equation}
where $v_1(K)$ the volume of a $d$-dimensional sphere of radius $K$ [cf. Eq. (\ref{v1})].
For $d=2$ and $d=3$, the ground states are disordered for $0 \le \chi < 1/2$.
At $\chi=1/2$, there is a  phase transition to a crystal phase.
The ``collective-coordinate" optimization procedure represents a powerful approach to 
generate numerically disordered stealthy hyperuniform many-particle systems \cite{Uc04b,Ba08,To15}.

A statistical-mechanical theory for disordered stealthy ground states has been formulated in the canonical ensemble
in the zero-temperature limit. By exploiting an ansatz that the entropically favored (most probable) stealthy
ground states in the canonical ensemble behave like ``pseudo" equilibrium  hard-sphere systems in Fourier space with an ``effective packing fraction"
$\phi$ that is proportional to $\chi$, one can employ well-established integral-equation formulations
for the pair correlation function of hard spheres in direct space \cite{Han13,To02a} to obtain
accurate theoretical predictions for $g_2({\bf r};\phi)$ and $S({\bf k};\chi)$ for a moderate range of $\chi$
about $\chi=0$ \cite{To15}, i.e.,
\begin{equation}
S({\bf k};\chi)=g_2^{HS}({\bf r=k};\phi).
\label{eq_sk}
\end{equation}
For example, the total correlation function $h(r)$ in the limit $\chi \rightarrow 0$ for any $d$, which
is exactly given by
\begin{equation}
\rho h(r) = -\left(\frac{K}{2 \pi r}\right)^{d/2}J_{d/2}(Kr) \qquad (\chi \rightarrow 0),
 \label{eq_hr}
 \end{equation}
corresponds to the following  unit-step function for the structure factor:
\begin{equation}
S({\bf k})=\Theta(k-K) \qquad (\chi \rightarrow 0).
\label{S-T-zero}
\end{equation}

Our interest in this paper are the excited states associated with the
long-ranged stealthy potentials \cite{To15} close to the ground states, i.e., when the absolute temperature $T$ is small. 
It was shown that the structure factor at the origin $S(0)$ varies linearly with $T$ for excited states.
Importantly, this positive value of $S(0)$ is the uniform
value of $S({\bf k})$ for $0 \le |{\bf k}|  \le K$ for the special case of the step-function
power-law potential for small $\chi$, which was  verified by
simulation results in various dimensions \cite{To15}. Because we are
interested in qualitative trends as the disordered ground state
is approached as the temperature is decreased, it suffices
for our purposes to consider the following simple form
for the structure factor at positive but small temperatures and sufficiently small $\chi$:
\begin{equation}
S({\bf k})= T^* \Theta(K-k)+ \Theta(k-K),
\label{S-excited}
\end{equation}
where $T^*$ is a dimensionless temperature that is much smaller than unity.
The form (\ref{S-excited}) is obtained by combining the aforementioned uniform value
for $0 \le |{\bf k}|  \le K$ with the ground state structure factor
as $\chi$ tends to zero, i.e., Eq. (\ref{S-T-zero}).

\begin{figure}[H]
\centering{\includegraphics[width = 0.5\textwidth,clip=]{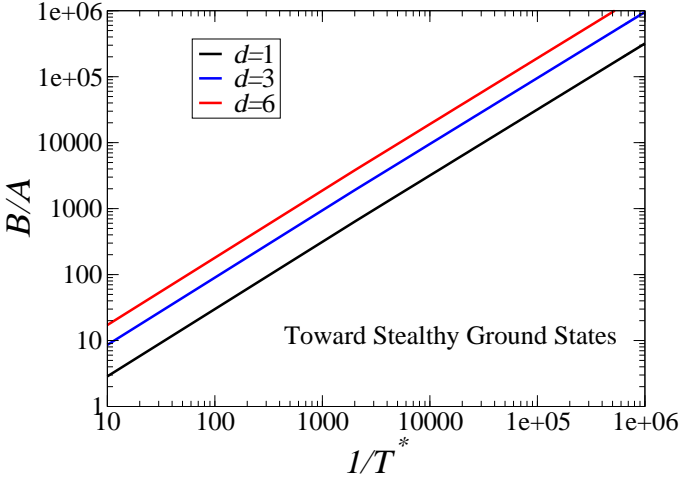}}
\caption{The ratio of the  surface-area coefficient to volume coefficient, $B/A$, as
a function of the inverse of the dimensionless temperature $1/T^*$ for excited
states associated with the long-ranged stealthy pair potential for dimensions $d=1,3$ and 6.}
\label{A-B-stealthy}
\end{figure}

It follows from (\ref{A}) and (\ref{S-excited}) that the volume coefficient
is given by 
\begin{equation}
A=T^*
\end{equation}
Moreover, substitution of (\ref{S-excited}) into the Fourier representation of the
surface-area coefficient (\ref{Fourier-B}) yields 
\begin{equation}
B=\frac{d}{\pi K}(1-T^*)
\end{equation}
Thus,  ratio $B/A$ is given by 
\begin{equation}
\frac{B}{A}=\frac{d}{\pi K} \frac{(1-T^*)}{T^*},
\label{B-A-stealth}
\end{equation}
which is plotted in Fig. \ref{A-B-stealthy} as a function of the inverse
of $T^*$ for selected space dimensions. We see that at fixed temperature, the ratio $B/A$ increases with the space dimension.
From relation (\ref{S-excited}), we see that the peak value of the structure factor is trivially unity ($S(k_{p})=1$), and hence
the hyperuniformity index is simply linear in $T^*$, i.e.,
\begin{equation}
H=T^*.
\label{H-stealth}
\end{equation}

\begin{figure}[H]
\centering{\includegraphics[width = 0.5\textwidth,clip=]{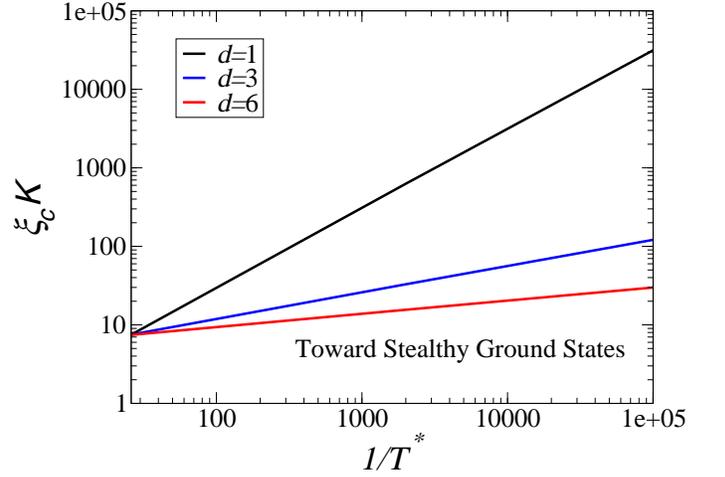}}
\caption{The dimensionless length scale $\xi_c K$ versus  the inverse of the dimensionless temperature $1/T^*$ for excited
states associated with the long-ranged stealthy pair potential for dimensions $d=1,3$ and 6.}
\label{length-scale-stealthy}
\end{figure}

Combination of (\ref{c}) and (\ref{S-excited}) yields the following expression
for the the Fourier transform of the direct correlation function for the excited states:
\begin{equation}
\rho {\tilde c}({\bf k})= (1-T^*) \Theta(K-k).
\end{equation}
Thus, evaluating this relation at zero wavenumber and use of (\ref{c}) yields
the length scale
\begin{equation}
\xi_c=(2\pi) \left[\frac{2d \, \chi}{v_1(K)}\right]^{1/d}\left(\frac{1}{T^*} -1\right)^{1/d}.
\label{xi_c-stealth}
\end{equation}
This length scale is plotted in Fig. \ref{length-scale-stealthy} as a function of the 
inverse temperature for selected space dimensions.

The three descriptors $B/A$, $H^{-1}$ and $\xi_c$ are increasing functions of
the inverse temperature.  From formulas  (\ref{B-A-stealth}) and (\ref{H-sticky}),
we observe that $B/A$ and $H^{-1}$  have the same scaling behavior as the inverse temperature
tends to infinity, i.e., the disordered  hyperuniform ground state is approached, namely,
\begin{equation}
\frac{B}{A} \sim H^{-1} \sim \frac{1}{T^*}.
\end{equation}
This is to be contrasted with the generally slower growth rate of $\xi_c$ with $T^*$
for $d \ge 2$, i.e.,
\begin{equation}
 (\xi_c)^d \sim \frac{B}{A} \sim H^{-1},
\end{equation}
where we have used relation (\ref{xi_c-stealth}).

\section{Conclusions and Discussion}

We have derived  a Fourier representation of the surface-area coefficient $B$ in terms of the structure factor $S({\bf k})$, which
is especially useful when scattering information is available experimentally or theoretically. 
In a disordered system that is nearly hyperuniform, we showed that the ratio of surface-area
to volume coefficients, $B/A$, enables one to ascertain hyperuniform and nonhyperuniform
distance-scaling regimes of the variance $\sigma^2(R)$ as a function of $R$
as well as the corresponding crossover distance between these regimes. 
While the analysis of the ratio $B/A$ was applied to systems that ultimately approached a class I
hyperuniform state, the corresponding extensions to other hyperuniformity classes is very straightforward.

Using the ratio $B/A$, as well as other diagnostic measures of hyperuniformity, including the hyperuniformity
index $H$ and the direct-correlation function length scale $\xi_c$, we characterized the structure of three different 
exactly solvable models as a function of the relevant control parameter, either
density or temperature, with end states that are perfectly hyperuniform. In the case of the
sticky hard-sphere and stealthy models, we studied the effect of dimensionality.
For all three models, we showed  that these diagnostic hyperuniformity measures are positively correlated with one another.
Thus, the quantities $H$ and $\xi_c$ can also be employed to infer the  hyperuniform and nonhyperuniform
distance-scaling regimes. This capacity to determine hyperuniform scaling regimes  is expected to be 
of great utility in analyzing experimentally- or computationally-generated samples that are necessarily of finite size.

In the same way that there is no perfect crystal,  due to the inevitable presence of imperfections, such
as vacancies and dislocations, there is no ``perfect” hyperuniform system in laboratory practice, 
whether it is ordered or not \cite{Ki18a}. It is clear that the same diagnostic measures explored in the present
work can be used to detect the degree to which such ``imperfect" nearly hyperuniform systems deviate from perfect
hyperuniformity. 

We showed that the free-volume theory of the pressure of
equilibrium packings of identical hard spheres that approach  a strictly jammed state, either along the stable crystal or metastable disordered branch,
dictates that such end states be exactly hyperuniform. This implies that the packing on approach to the jammed state
must be ergodic in a generalized  sense and hence free of any defects, e.g., vacancies or dislocations
along the crystal branch or rattlers along the metastable fluid branch. It is an outstanding problem for future research to
place these implications of free-volume theory for the hyperuniformity of strictly jammed packings on
a firmer rigorous theoretical foundation.



{\appendix

\section{Number Variance Scalings for Nonhyperuniform Systems}
\label{scalings}

The local number variance $\sigma^2_{_N}(R)$  is generally a function that can be decomposed
into a global part that grows with the window radius $R$ and a local part that oscillates on
small length scales  (e.g., mean nearest-neighbor distance) about the global contribution.
The more general large-$R$ asymptotic formula for the variance is given by \cite{To18a}
\begin{equation}
\sigma^2(R)=
2^d\phi\Bigg[ A_{_N}(R)\left(\frac{R}{D}\right)^d+B_{_N}(R)\left(\frac{R}{D}\right)^{d-1}+
o\left(\frac{R}{D}\right)^{d-1}\Bigg],
\label{var1}
\end{equation}
where the $R$-dependent volume and surface-area coefficients   are respectively given by
\begin{equation}
A_{_N}(R) =1+\frac{\phi}{v_1(D/2)} 
\int_{|{\bf r}| \le 2R} h({\bf r}) d{\bf r}
\label{A1} 
\end{equation}
and
\begin{equation}
B_{_N}(R)=-\frac{\phi\,d\, \Gamma(d/2)}{2\,\pi^{1/2} ,D\,v_1(D/2)\Gamma[(d+1)/2]} \int_{|{\bf r}| \le 2R} h({\bf r})|{\bf r}| d{\bf r}.
\label{B1}
\end{equation}
Observe that when the volume coefficient $A_{N}(R)$ and surface-area coefficient  $B_{N}(R)$
 converge in the limit $R \rightarrow \infty$, they are  equal
to the constants $A$ and $B$, defined by ({\ref{A}) and (\ref{B}), respectively. For this reason, $A$ and $B$
are called the {\it global} volume and surface-area coefficients, respectively \cite{To18a}.

An anti-hyperuniform system is one in which the exponent $\alpha$ [cf. (\ref{eq:Sk-scaling})] is negative $\alpha<0$,
resulting in a structure factor that diverges in the zero-wavenumber limit.
The corresponding total correlation function $h({\bf r})$ decays like $1/r^{d +\alpha}$ for
large $r\equiv |\bf r|$. This power-law tail controls the growth rate of $A_{_N}(R)$ with $R$; specifically,
carrying out the integration in (\ref{A1}) yields 
\begin{equation}
A_{_N}(R) \sim R^{-\alpha}.
\end{equation}
Substitution of this scaling behavior in the leading-order term of (\ref{var1}) gives
that the local number variance for an anti-hyperuniform system scales like $\sigma^2(R) \sim R^{d-\alpha}$,
which proves the anti-hyperuniform scaling given in (\ref{sigma-nonhyper}).
Since $\alpha$ is negative, the number variance grows faster than
the window volume, i.e., faster than $R^d$. Up to a trivial constant in the leading-order
term in the asymptotic expansion of the number variance given by (\ref{var1}), one can
view a ``typical" nonhyperuniform system in which $S(0)$ is bounded as
one in which $\alpha=0^+$ (approaches zero from above), even if $S(\bf k)$
is not described by a power-law form in the zero-wavenumber limit.  This enables
one to conclude that $A_{_N}(R)$ converges to a constant in the limit $R \rightarrow \infty$,
thus yielding the $R^d$ scaling behavior for $\sigma^2(R)$ in (\ref{sigma-nonhyper}).

\section{Evaluation of the coefficients $A$ and $B$ for a Super-Poissonian Point Pattern}
\label{super}

Any nonhyperuniform point process for which $S(0) >1$ has a large-$R$ asymptotic number variance $\sigma^2(R)$ that is larger
than that for a Poisson point process [$S(0)=1$] with the same mean $\langle N(R)\rangle$ is
called {\it super-Poissonian} \cite{To21b}.
We evaluate the volume and surface-area coefficients $A$ and $B$ for a specific model
of a super-Poissonian point process, namely, the  \textit{Poisson cluster process},
which is characterized by strong clustering of the points 
with a large but finite value of $S(0)$, and hence  is far from being hyperuniform. The higher-order moments
of the number fluctuations as well as the corresponding probability distribution for this model
were recently studied \cite{To21b}. The construction of the cluster process starts from a homogeneous
Poisson point process of intensity  $\rho_p$~\cite{La17}.
Each point of the Poisson point process is the center of a cluster of points. The
number of points in each cluster is independent and follows a
Poisson distribution with mean value $c$. Following Ref. \cite{To21b}, we consider
here the special case in which the positions of the points relative to the
center of the cluster follow an isotropic Gaussian distribution with
standard deviation  $r_0$, which can be regarded to be the characteristic length scale of a single cluster.
In the infinite-volume limit, the  structure factor  for any $d$ is exactly given by \cite{To21b}
\begin{equation}
S(k)= 1+ c e^{- k^2 r_0^2}.
\label{S-super}
\end{equation}
It immediately follows from this formula and (\ref{A}) that $S(0)=A=1+c$. Using the Fourier representation of the surface-area
coefficient (\ref{Fourier-B}) and relation (\ref{S-super}), we find that $B$, for any $d$, is exactly given by
\begin{equation}
B= - \frac{c\, r_0\, d}{\sqrt{\pi}}.
\label{B-super}
\end{equation}
Remarkably, this model provides an example of a surface-area coefficient that is negative (when $c$ is positive), which
heretofore has not been identified. Since the surface-area coefficient $B$ is derived from the large-$R$ asymptotic
expansion (\ref{eq:sigma}) of the number variance $\sigma^2(R)$, which must be positive, the length scale $r_0$ in (\ref{B-super}),
as any characteristic length scale, must be much smaller
than the window radius $R$.

\section{Sum Rules and Large-$k$ Asymptotic Behaviors of $S(k)$ for Sphere Packings}
\label{sum-rule}

Here we present sum rules as well as the  exact large-$k$ asymptotic behaviors of the structure
factors of certain general packings of identical spheres that apply in any space dimension $d$.

For a large class of packings of identical spheres of diameter $D$, the following sum rule
applies inside the hard core:
\begin{equation}
\frac{r}{\left(2\pi r\right)^{\frac{d}{2}}}\int_{0}^{\infty}k^{d/2} {\tilde h}(k) J_{\left(d/2\right)-1}(k r)\, dk=-1 \qquad \mbox{for}\;\;r <D.
\label{inverse-1}
\end{equation}
This sum rule follows from the fact that $h({\bf r})=-1$ for $r <D$ 
and use of (\ref{Inverse}). It is valid provided that the volume integral of ${\tilde h}({\bf k})$ over all reciprocal space is bounded. 
In the special case of $r=0$, (\ref{inverse-1}) yields
\begin{equation}
\frac{d }{2^{d} \pi^{d/2}\Gamma(1+d/2)} \int_{0}^{\infty}k^{d-1} {\tilde h}(k) \, dk=-1.
\label{inverse-2}
\end{equation}
Such exact sum rules can be utilized to check the accuracy of
the determination of $S(k)$ of sphere packings  via experimental or numerical methods.

The Fourier transform of the indicator function $m(r;a)=\Theta(a-r)$ for a $d$-dimensional sphere of radius $a$ is given by
\begin{eqnarray}
{\tilde m}(k;a) &=& \frac{(2\pi)^{d/2}}{k^{(d/2)-1}}\int_0^a
r^{d/2} J_{(d/2)-1}(kr) dr \nonumber \\
&=& \left(\frac{2\pi}{ka}\right)^{d/2} a^d J_{d/2}(ka).
\label{window-Fourier}
\end{eqnarray}
For large $k$, this Fourier transform is given by
\begin{equation}
{\tilde m}(k;a) \sim R^d (2\pi)^{d/2} \sqrt{2/\pi}\; \frac{\cos[ka-(d+1)/4]}{(ka)^{(d+1)/2}}+{\cal O}\left(\frac{1}{(ka)^{(d+3)/2}}\right) \qquad (ka \to \infty).
\label{m}
\end{equation}

For any packing of spheres of diameter $D$, $h(r)=-m(r;D)$ for $r < D$. Let us assume that the contact-value of the pair correlation function, denoted by $g_2(D^+)$, is bounded, i.e., the jump discontinuity at contact is finite. Using the result (\ref{m}) for such a packing, we can relate the large-$k$ behavior of the structure factor  in terms
of $g_2(D^+)$:
\begin{eqnarray}
S(k) &&\sim 1-  \phi \, g_2(D^+)\, 2^{3d/2} \Gamma(1+d/2)  \sqrt{2/\pi}\; \frac{\cos[kD-(d+1)/4]}{(kD)^{(d+1)/2}}
\nonumber \\
&&\hspace{0.35in}+{\cal O}\left(\frac{1}{(kD)^{(d+3)/2}}\right)\qquad (kD \to \infty).
\end{eqnarray}
For $d=1,2$ and 3, the corresponding  asymptotic behaviors are explicitly given respectively by
\begin{equation}
S(k) \sim 1-  2 \phi \, g_2(D^+) \, \frac{\sin(kD)}{kD}+{\cal O}\left(\frac{1}{(kD)^{2}}\right) \quad (d=1),
\label{S-largek-1}
\end{equation}
\begin{equation}
S(k) \sim 1-  8  \sqrt{2/\pi}\, \phi \, g_2(D^+) \, \frac{\cos(kD+\pi/4)}{(kD)^{3/2}}+{\cal O}\left(\frac{1}{(kD)^{5/2}}\right) \quad (d=2),
\end{equation}
and
\begin{equation}
S(k) \sim 1+  24 \phi \, g_2(D^+)\, \frac{\cos(kD)}{(kD)^2}+{\cal O}\left(\frac{1}{(kD)^{3}}\right) \quad (d=3).
\end{equation}
We see that the rate of decay of $S(k)$ increases as $d$ increases.

By contrast, if the sphere packing is disordered at a jammed state with average
contact number per particle of $Z$, then the jamming-contact condition
is described by
\begin{equation}
g_2(r) \sim \frac{Z}{\rho s_1(1) D^{d-1}}\delta(r-D),
\end{equation}
where $s_1(1)=d v_1(1)$ is the $d$-dimensional surface area
of a sphere of unit radius. Thus, we see for such jammed packings, the large-$k$ behavior of the structure factor
is given by
\begin{eqnarray}
S(k) &\sim& 1 + \frac{2^{d/2} \Gamma(1+d/2)\, Z}{d} \,\frac{J_{d/2 -1}(kD)}{(kD)^{d/2-1}} \nonumber \\
&& +{\cal O}\left(\frac{1}{(kD)^{(d+1)/2}}\right) \qquad (kD \to \infty).
\label{S-largek}
\end{eqnarray}
For $d=1,2$ and 3, the corresponding  asymptotic behaviors are explicitly given respectively by
\begin{equation}
S(k) \sim 1 +  Z \,\cos(kD) + {\cal O}\left(\frac{1}{(kD)}\right)\qquad (d=1)
\end{equation}
\begin{equation}
S(k) \sim 1  + Z\, J_0(kD) +{\cal O}\left(\frac{1}{(kD)^{3/2}}\right) \qquad (d=2)
\end{equation}
\begin{equation}
S(k) \sim 1  +  Z \; \frac{\sin(kD)}{kD} + {\cal O}\left(\frac{1}{(kD)^{2}}\right)\qquad (d=3)\,.
\end{equation}
Again, we see that the rate of decay of $S(k)$ increases as $d$ increases, but less rapidly
than in unjammed packings at the same dimension in which the jump discontinuity in $g_2(r)$
at contact is finite.

\begin{acknowledgements}
The author is grateful to Jaeuk Kim, Michael Klatt, Charles Maher, Murray Skolnick and Haina Wang 
for their careful reading of the manuscript. This work was supported  by the National Science Foundation
under Award No. DGE-2039656.
\end{acknowledgements}



%

\end{document}